# Testing local position and fundamental constant invariance due to periodic gravitation and boost using long-term comparison of the SYRTE atomic fountains and H-masers

M. E. Tobar, P. L. Stanwix, J.J. McFerran

School of Physics, University of Western Australia, 6009, Crawley, Australia

michael.tobar@uwa.edu.au

J. Guéna, M. Abgrall, S. Bize, A. Clairon, Ph. Laurent, P. Rosenbusch, D. Rovera, G. Santarelli

LNE-SYRTE, Observatoire de Paris, CNRS, UPMC 75014, Paris, France

The frequencies of three separate Cs fountain clocks and one Rb fountain clock have been compared to various hydrogen masers to search for periodic changes correlated with the changing solar gravitational potential at the Earth and boost with respect to the Cosmic Microwave Background (CMB) rest frame. The data sets span over more than eight years. The main sources of long-term noise in such experiments are the offsets and linear drifts associated with the various H-masers. The drift can vary from nearly immeasurable to as high as $1.3\times10^{-15}$ per day. To circumvent these effects we apply a numerical derivative to the data, which significantly reduces the standard error when searching for periodic signals. We determine a standard error for the putative Local Position Invariance (LPI) coefficient with respect to gravity for a Cs-Fountain H-maser comparison of $|\beta_H - \beta_{Cs}| \leq 4.8\times10^{-6}$ and $|\beta_H - \beta_{Rb}| \leq 10^{-5}$ for a Rb-Fountain H-maser comparison. From the same data the putative boost LPI coefficients were measured to a precision of up to parts in $10^{11}$ with respect to the CMB rest frame. By combining these boost invariance experiments to a Cryogenic Sapphire Oscillator versus H-maser comparison[1], independent limits on all nine coefficients of the boost violation vector with respect to fundamental constant invariance, $\boldsymbol{B}_\alpha$, $\boldsymbol{B}_e$ and $\boldsymbol{B}_q$ (fine structure constant, electron mass and quark mass respectively), were determined to a precision of parts up to $10^{10}$.

PAC numbers: 04.80.Cc, 03.30.+p, 06.30.Ft

I. **INTRODUCTION**

Local Position Invariance (LPI) and Local Lorentz Invariance (LLI) are fundamental components of the theory of General Relativity (GR) and the Einstein Equivalence Principle (EEP) [2,3]. GR is a major scientific theory whose validity has significant implications for frontier research and every day applications (such as GPS). It is therefore appropriate to experimentally validate the basic tenets of GR as precisely as possible. The precision of this validation in turn determines the level of confidence to which theoretical predictions can be deployed. Furthermore, a number of unification theories suggest violations of EEP at some level [4-7], which enhances the motivation for these types of tests. However, to identify violations it is necessary to have an alternative test theory to interpret the experiment, which have been developed for both LLI experiments [5, 6, 8-11] and LPI experiments [2, 4, 7, 12-14]. It is the parameters of these test theories than can be experimentally tested to search for new physics beyond GR.

Specifically for this work we test LPI; the principle of LPI is that the outcome of any local non-gravitational experiment is independent of where and when it is performed. Such experiments test for spatial and temporal invariance of fundamental physics. Traditionally, spatial LPI is tested through gravitational redshift experiments [15-24], which compare the frequency of the radiation emitted from two different clocks or oscillators at the same position. The invariance of LPI dictates that the gravitational redshift of the clocks (or oscillators) frequency is universal and independent on clock type. However, the gravitational field is not the only changing physical parameter that could affect a clock, and such experiments could be broadened to test against other temporally or spatially varying physical parameters. For example, varying boost with respect to a chosen frame of reference could in principle cause a similar spatial dependence of clock frequencies if new physics exists [1]. In the case of boost, the invariance of LPI is associated with the Doppler-shift of the clocks or oscillators under comparison. Thus, in analogy to a "null redshift" experiment, a "null Doppler-shift" experiment could be performed between two different clocks in the same location and with a common varying boost with respect to a chosen frame of reference.

Modern LPI tests make use of detailed calculations of the frequency dependence of atomic transitions on fundamental constants, which show that each atomic species has its own unique dependence [1, 12, 25]. This gives a mechanism for measuring a putative frequency variation between two different clocks. Also, measurements of quasar spectra over cosmological redshifts suggest that fundamental constants may vary over cosmic scales [26-32]. Furthermore, it has been shown that varying coupling constants, such as the fine structure constant, can be modelled as a violation of Lorentz and CPT symmetry [33]. These reasons give us incentive to also use clocks to undertake

Doppler-shift LPI experiments due to boost with respect to a cosmic frame of reference, testing for a null hypothesis. The logical choice of reference frame to test against is the Cosmic Microwave Background (CMB), as it is a cosmic frame of reference and has already been chosen for testing LLI, such as Kennedy-Thorndike, Michelson-Morley and Ives-Stillwell experiments [34-36].

More modern tests of LLI use a general Lorentz violating extension of the standard model of particle physics (SME), whose Lagrangian includes all parameterized Lorentz violating terms that can be formed from known fields [5, 6, 37, 38]. This has inspired a new wave of experiments designed to explore uncharted regions of the SME Lorentz violating parameter space (see [39] and references therein for a comprehensive list of all parameters and precision of experimental constraints). These experiments test new physics with regards to boost with respect to a Sun Centered Frame, which remains inertial with respect to the CMB. The experiment reported here may be sensitive to some of these parameters, however testing LLI is beyond the scope of this paper. The focus of this work is to undertake a new LPI test of boost (null Doppler shift experiment) with respect to the CMB. Thus, we are able to place limits on the variation of fundamental constants with respect to the solar gravitational potential on Earth and boost with respect to the CMB rest frame. Furthermore, by combining the experiments in this work to that in reference [1], we can perform a first determination of all nine coefficients of the boost violation vectors $\boldsymbol{B}_\alpha$, $\boldsymbol{B}_e$, and $\boldsymbol{B}_q$, with respect to fine structure constant, electron mass and quark mass respectively.

## II. CLOCK COMPARISON

The LNE-SYRTE clock ensemble at the Paris Observatory combines an array of high precision atomic clocks and oscillators at microwave and optical frequencies, and has been described in detail elsewhere, see for example the recent review paper by [40]. In this work we concentrate on the long-term comparison between a number of H-masers, three Cs fountains and one Rb fountain at LNE-SYRTE. All transitions involved are hyperfine structure transitions and at a microwave frequency. Data was typically recorded every 100 seconds, thus producing a large amount of data after several years of comparisons. Since the periods of the signals of interest are much longer than the sampling interval, the data was averaged over longer periods to reduce the amount of data for subsequent processing. This is especially important when implementing computationally intensive noise optimizing algorithms like weighted least squares (WLS), which whitens the noise and reduces the effects of the noise in the search for small signals. To search for annual variations the data sets are averaged over 95,000 sec intervals (1.1 days) and to search for sidereal variations the data sets are averaged over 2,500 second intervals (42 minutes).

Figure 1 shows the directly measured fractional frequency offset variations of the frequency ratio between the H-masers and the fountain clocks at the Paris Observatory ($x_{Cs-H} = \nu_{Cs}/\nu_H$ and $x_{Rb-H} = \nu_{Rb}/\nu_H$). The dominant noise source masking any periodic variation is due to offsets and drifts generated by the H-masers. This is why in general all three Cs comparisons overlap one another as a function of time, as the H-maser used in the comparison is common to all three fountains. It has been show that there is an advantage to analysing the derivative of the beat frequency over the beat frequency directly, because it naturally filters out non-stationary effects such as systematic jumps and drifts, leading to a more sensitive measurement [1]. The technique converts frequency jumps and other systematic offsets to outliers. In general if drifts are small they manifest as relatively small offsets compared to the remaining statistical noise. The quality of the data can be further improved when the length of a subset of continuous data is greater than the period of interest (here annual and sidereal periods). In this case the drift may be removed from the original data, which in turn further eliminates some of the small offsets that occur after taking the derivative. This technique has been implemented for both sets of data in figure 1, and is presented in figure 2 and 3 for the annual and sidereal time scales respectively.

Frequency measurements can only be made through a comparison that utilizes two or more clocks simultaneously. Therefore, a key point is that only when two clocks have different sensitivities to fundamental constants can a putative LPI violation be apparent. In general all such clock comparisons measure a fractional frequency ratio, $x_{1-2}$, between the clocks (labelled 1 and 2), which takes the form [41]:

$$x_{1-2} = \frac{\nu_1}{\nu_2} = Const \times \alpha^{n_\alpha} \mu_e^{n_e} \mu_q^{n_q} \: . \tag{1}$$

Here $\alpha$, is the fine structure constant (or electromagnetic coupling constant), $\mu_q = m_q/\Lambda_{QCD}$, where $m_q$ is the quark mass, $\Lambda_{QCD}$ the mass scale of the strong force in quantum chromodynamics (QCD), and the electron-to-proton mass ratio $\mu_e = m_e/m_p$ (or $m_e/\Lambda_{QCD}$). In the case of the Cs versus H-maser comparison, $n_\alpha = 0.83$, $n_e = 0$ and $n_q = 0.102$; for the Rb versus H-maser comparison $n_\alpha = 0.34$, $n_e = 0$ and $n_q = 0.081$ [41, 42]. Because the clock comparisons defined by (1) depend on three dimensionless fundamental constants, one needs to undertake at least three separate measurements of three pairs of clocks with different values of $n_\alpha$, $n_e$ and $n_q$ to be able to uniquely determine any supposed LPI effects.

### III. SEARCH FOR PERIODIC VARIATIONS

In this section we summarize the results of searching for periodic signals in the Cs versus H and Rb versus H clock comparisons. The whole process from measurement to final analysis is a six-step process as shown in figure 4. The first three steps were presented in detail in the prior section, which describes the initial manipulation to prepare the data before algorithms are applied to search for periodic signals, which potentially indicate new physics. The last three steps are detailed in this section, which includes analysis of the data to optimize the search using the Weighted Least Squares (WLS) technique.

Previous clock-comparison experiments that searched for periodic signals have mainly tested LPI as a "null redshift" experiment. Recent experiments include Ashby et al. [20], who achieved 1.4 ppm comparing H-Cs, Blatt et al. [18] who achieved 3.5 ppm comparing Cs-Sr, Fortier et al. [21] who achieved 3.5 ppm comparing Hg-Cs, and Guena et al. [23] who achieved 1.0 ppm using Cs-Rb. The work presented here has used the same Cs and Rb fountains implemented in Guena et al., but measured independently against various H-masers. The H-masers in this work reduce the sensitivity of the comparisons significantly, to 10 ppm for the Rb-H maser comparison and 4.8 ppm for the Cs-H maser comparison (see next section). However, comparing the clocks in this way generates three independent clock comparisons, enabling the first de-correlation of the boost violation vector with respect to fundamental constant invariance [1]. Furthermore, the work represents only the second Rb-H comparisons in the literature, which was only just recently published by Peil et al. [24] in 2013 achieving 0.5 ppm.

#### A. Search for LPI variations with respect to gravitational potential of the Sun

The gravitational potential at the Earth varies due to its elliptic orbit around the Sun at a period equal to its orbital period. Thus, data of figure 2 can be used to search for a signal of the form $-\Omega_\oplus C_{\Omega_\oplus} Sin[\Omega_\oplus(t-t_o)]$. Here $C_{\Omega_\oplus}$ is the quadrature amplitude in-phase with the changing solar gravitational potential on Earth at the annual angular frequency, $\Omega_\oplus$, with $t_0$ setting the phase to zero at one of the perihelion's (which in this case is on the 4th of January 2003). To optimize the search WLS is implemented, with the weighting function determined by taking the power spectral density of the data in fig. 2, then fitting the spectral density to a power law fit near the Fourier frequency of interest. The noise is then whitened around this frequency, details of the process is similar to that in [1]. Figure 5 shows the simultaneous fit for the in-phase and out-of-phase amplitudes at the annual frequency and twice the annual frequency. At the annual frequency these fits do not show any offsets from zero beyond two standard errors of the fit. However, at twice the annual frequency the out of phase component is close to thee standard errors in amplitude for the Cs versus H-maser comparison, which may be the sign of the beginning of some sort systematic effect at this frequency.

For this type of test, the amplitude of the LPI violation is given by [17-19] [20] [21]:

$$C_{\Omega_\oplus LPI} = -(\beta_1 - \beta_2)\frac{Gm_s}{ac^2}e \qquad (2)$$

Here $\beta_1$ and $\beta_2$ are the putative gravitational violation coefficients for the two clocks in the comparison, $G$ is the gravitational constant, $c$ the speed of light, $m_s$ the Sun's mass, $a$ the semi major axis of the Earth's orbit, $e$ the eccentricity, where we have $-eGm_s/ac^2 \approx -1.65 \times 10^{-10}$. The value for $C_{\Omega_\oplus LPI}$ for the two clock comparisons is presented in figure 5, and is determined to be $C_{\Omega_\oplus LPI}$ = 5.9(7.9)×10$^{-16}$ for the Cs versus H comparison and $C_{\Omega_\oplus LPI}$ = 1.0(2.1)×10$^{-15}$ for the Rb versus H comparison. Thus from (2) we determine $|\beta_{Rb} - \beta_H|$ = 6.3(10) ×10$^{-6}$ and $|\beta_{Cs} - \beta_H|$ = 3.6(4.8) ×10$^{-6}$.

A very recent publication shows that the current best test for the Rb versus H is 4.9×10$^{-7}$ and Cs versus H is 1.1×10$^{-6}$, so here we are 20 times and 4.4 times less sensitive respectively [17]. However, this is only the second measurement of the former, and the later is of the same order as the current best test. Also, the data in [17] spans only 1.5 years, in contrast our comparison, which is over 8 years and is more significant in the search for an annual modulation. Another difference is that the result in [17] relied on a large ensemble of H-masers and fountains with high stability but was not evaluated in accuracy. In our case the fountains are few (only one in case of the Rb versus H-maser comparison) and are not only stable but also are also accurate. Even though these results are not the best, they are important to present here as the coefficients determined from the "null redshift" analysis may be compared to the "null Doppler shift" analysis presented in the next section as they are determined from exactly the same sets of data. It should be noted that the many experiments that have been used to determine "null redshift" coefficients could be used to do sensitive "null Doppler shift" experiments by following the same approach.

It is usual to use the "null redshift" experiments to determine possible variation of fundamental constants with respect to gravitational field. Thus from (1), it is usual to define the following sensitivity coefficients:

$$\kappa_\alpha = \frac{\delta\alpha/\alpha}{\delta\left(\frac{GM}{rc^2}\right)}; \quad \kappa_e = \frac{\delta\mu_e/\mu_e}{\delta\left(\frac{GM}{rc^2}\right)}; \quad \kappa_q = \frac{\delta\mu_q/\mu_q}{\delta\left(\frac{GM}{rc^2}\right)} \qquad (3)$$

Then from the known values of $n_\alpha$, $n_e$ and $n_q$, we can obtain a limit on these values of $0.83\kappa_\alpha + 0.102\kappa_q = 3.6(4.8) \times 10^{-6}$ and $0.34\kappa_\alpha + 0.081\kappa_q = 6.3(10) \times 10^{-6}$ from the Cs versus H and Rb versus H comparisons respectively. Likewise in the next section we take the same approach and determine coefficients of possible fundamental constant variation with respect to boost.

B. **Search for LPI variations with respect to boost with respect to the CMB**

To put a limit on variation with respect to boost (null Doppler shift experiment), we use an Earth cantered frame as defined in [1][43][44] as shown in figure 6. Assuming a LPI violation, which is non-zero with respect to boost and of the form shown in (1), we can define a frequency shift of the form [1]:

$$\frac{\delta x_{1-2}}{x_{1-2}} = \mathbf{B} \cdot \mathbf{b}; \quad \mathbf{B} = n_\alpha \mathbf{B}_\alpha + n_e \mathbf{B}_e + n_q \mathbf{B}_q \tag{4}$$

Where

$$\mathrm{B}_{\alpha\,i} = \frac{1}{\alpha}\frac{\delta\alpha}{\delta b_i}; \quad \mathrm{B}_{e\,i} = \frac{1}{\mu_e}\frac{\delta\mu_e}{\delta b_i}; \quad \mathrm{B}_{q\,i} = \frac{1}{\mu_q}\frac{\delta\mu_q}{\delta b_i} \tag{5}$$

Here $\mathbf{b}$ is the boost vector of the local position of the experiment with respect to the CMB, and the putative violation due to the boost is described by the vector $\mathbf{B} = B_x\hat{x} + B_y\hat{y} + B_z\hat{z}$ with respect to the Earth Frame in figure 6, which is a function of the vectors $\mathbf{B}_\alpha$, $\mathbf{B}_e$ and $\mathbf{B}_q$ with components $i = x, y$ or $z$ defined in (5). Thus, equation (5) defines nine components to put limits on. To succeed in limiting all nine components, we need three separate combinations of clocks to compare, which have different values of $n_\alpha, n_e$ and $n_q$ given in (4), with a long enough set of data to fit for both sidereal and annual frequencies. This has been achieved by combining the results in this work (Rb versus H and Cs versus H comparison) with [1] (CSO[45] versus H comparison). For the least squares analysis over the annual period only the phase of the quadratures of figure 5 needs to be changed to be with respect to the CMB rather than the gravitational potential. Thus, the methods used and the statistics of the results are very similar. Similarly, to fit for the sidereal component of $\mathbf{B}$ from the data in figure 3, the quadratures are set by the phase of the boost with respect to the CMB rest frame as in [1], with the quadrature amplitudes and standard errors shown in figure 7 for a variety of frequencies near the sidereal period.

The results of fitting for boost components to put limits on $\mathbf{B}$ are tabulated in Table I, which presents the fits to the sidereal and annual components and their relationship to the linear combination of components of the boost violation vector with respect to the CMB. The system is over constrained, so we use least squares to determine the individual components. The result of this analysis is presented in Table II for the three separate clock comparisons. For the H-maser -

CSO comparison it was shown previously that $\boldsymbol{B}_{H-CSO} = 3\boldsymbol{B}_\alpha + \boldsymbol{B}_e - 0.1\boldsymbol{B}_q$ so that $n_\alpha = 3$, $n_e = 1$ and $n_q = -0.1$ [1]. Thus, including this data into Table II gives nine measurements so the nine coefficients of the boost violation vector with respect to fundamental constant invariance, $\boldsymbol{B}_\alpha, \boldsymbol{B}_e$ and $\boldsymbol{B}_q$ can have individual limits set on their values, which is presented in Table III.

Table I. From the clock comparison data of figures 2 and 3, and assuming the frequency deviation is given by (4). The in-phase and quadrature components of the annual ($\Omega_\oplus$) and sidereal ($\omega_\oplus$) components (shown in column 1) have amplitudes dependent on the components of the boost violation vector $\boldsymbol{B} = B_x\hat{x} + B_y\hat{y} + B_z\hat{z}$ (shown in column 2) [1] for the three separate clock comparisons.

| Frequency Component | Component Amplitude | CSO / H-maser [1] | Cs / H-maser | Rb / H-maser |
|---|---|---|---|---|
| $\sin[\Omega_\oplus(t-t_o)]$ | $9.71\times10^{-5}B_x + 2.08\times10^{-5}B_y$ | -5.4 (2.4) ×10$^{-14}$ | -2.9(8.0)×10$^{-16}$ | -3.4 (1.9)×10$^{-15}$ |
| $\cos[\Omega_\oplus(t-t_o)]$ | $1.91\times10^{-5}B_x - 8.92\times10^{-5}B_y + 3.92\times10^{-5}B_z$ | -2.7 (2.1) ×10$^{-14}$ | 11(7.8)×10$^{-16}$ | -1.9 (2.2)×10$^{-15}$ |
| $\sin[\omega_\oplus(t-t_o)+\Phi]$ | $-1.02\times10^{-6}B_x$ | 5.6 (10) ×10$^{-17}$ | -8.4(4.3)×10$^{-17}$ | -18(8.2)×10$^{-17}$ |
| $\cos[\omega_\oplus(t-t_o)+\Phi]$ | $1.02\times10^{-6}B_y$ | 8.7 (10) ×10$^{-17}$ | -6.3(4.3)×10$^{-17}$ | 3.5(8.2)×10$^{-17}$ |

Table II. The data in Table I is over constrained by the measurements. Thus, least squares analysis was implemented to determine each component of the boost violation vector with associated standard error.

| Boost Violation Vector Component | H-maser / CSO [1] | Cs / H-maser | Rb / H-maser |
|---|---|---|---|
| $B_x$ | -12.6 (9.1) ×10$^{-11}$ | 1.6(1.1)×10$^{-11}$ | -2.2 (2.6)×10$^{-11}$ |
| $B_y$ | 6.5 (9.8) ×10$^{-11}$ | -7.6(3.3)×10$^{-11}$ | -0.8 (3.6)×10$^{-11}$ |
| $B_z$ | -46.8 (48.5) ×10$^{-11}$ | -15(15)×10$^{-11}$ | -5.6 (9.7)×10$^{-11}$ |

Table III. Decomposition of the boost violation vectors in Table II to limits on the invariance of fundamental constants with respect to boost. The values range from 10$^{-10}$ to a few parts in 10$^{-9}$.

| Fundamental Constant Boost Violation Vector | $i = x$ | $i = y$ | $i = z$ |
|---|---|---|---|
| $B_{\alpha i}$ | 1.1 (0.9) ×10$^{-10}$ | -1.6(1.4)×10$^{-10}$ | -2.0 (4.8)×10$^{-10}$ |
| $B_{e i}$ | -5.3 (3.4) ×10$^{-10}$ | 6.2(5.2)×10$^{-10}$ | 1.4 (18)×10$^{-10}$ |
| $B_{q i}$ | -7.3 (6.7) ×10$^{-10}$ | 5.9(9.8)×10$^{-10}$ | 1.4 (29)×10$^{-10}$ |

Because the violations are measured with respect to an Earth cantered frame the sidereal components do not couple to the z-component of the boost violation vector. This means the z-components as presented in Table II and III inevitably have a weaker constraint as this term is only coupled through the annual components.

## IV. Discussion and Conclusions

In this work a major source of known non-stationary noise is eliminated by numerically differentiating the data. This source largely comes from the H-masers and not the fountains, due to unknown offsets and jumps in the frequency. The resulting measurements here show no significant departure from zero within the standard error estimates, which indicates effects of unknown systematic errors are most likely minimal, or at least of the order of the standard errors presented. Due to the very low frequencies of interest in the analysis (sidereal and annual) it is difficult to completely determine the exact effects of systematic errors on the data. However, the approach presented here only requires a measurement of the relative stability and not an absolute measurement of the frequency. Thus only systematic errors that correlate with the frequencies of interest are important. In other words, constant systematic shifts or shifts at frequencies other than the frequency of interest, are not important. Since we determine null measurements from sidereal and annual frequencies, this indicates that systematic errors are not limiting our experiment at the frequencies of interest. This has been discussed in detail in [1] for the CSO – H-maser comparison. For the Cs-H and Rb-H comparisons shown here, if a systematic was present one would expect larger signals at the annual and sidereal frequencies compared to other frequencies, which are fitted using WLS, with results shown in figures 5 and 7. No unusual large amplitudes were seen leading us to conclude that there are only small systematic influences, which has lead to a null result with the LPI coefficients with respect to gravity measured to be $|\beta_H-\beta_{Cs}| \leq 4.8 \times 10^{-6}$ and $|\beta_H-\beta_{Rb}| \leq 10^{-5}$. From the same data boost LPI coefficients were measured to a precision of up to parts in $10^{11}$ with respect to the CMB rest frame. For the first time independent limits on all nine coefficients of the boost violation vector with respect to fundamental constant invariance $\boldsymbol{B}_\alpha$, $\boldsymbol{B}_e$ and $\boldsymbol{B}_q$ were determined to a precision of up to parts in $10^{10}$.

**Figure Captions**

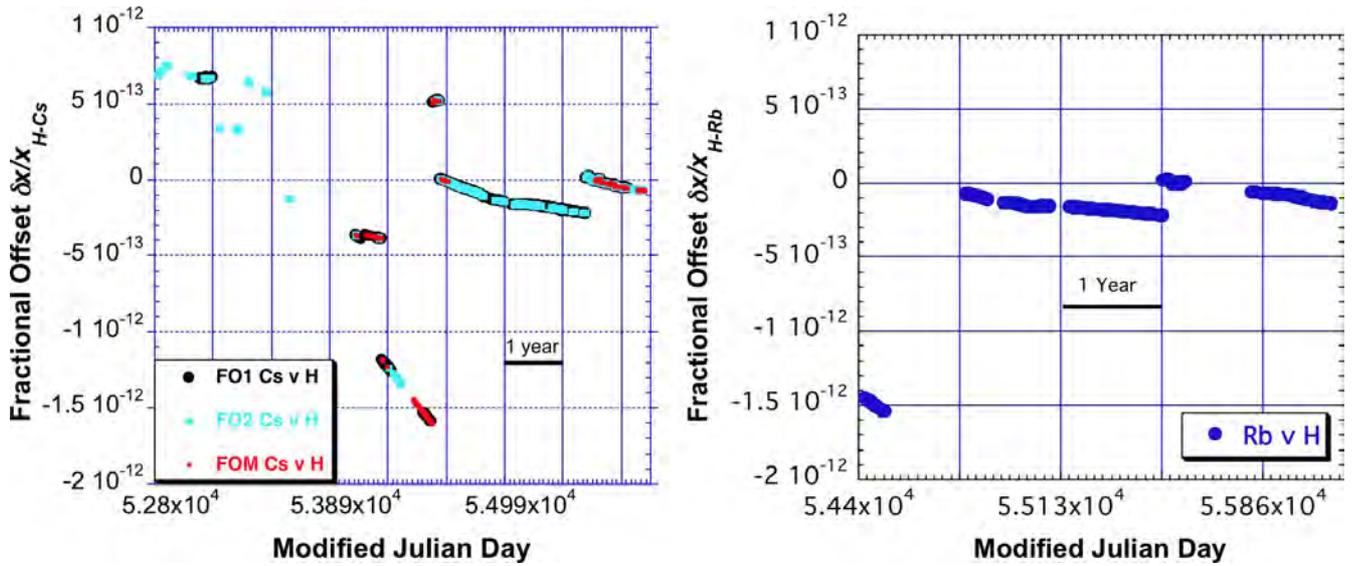

Figure 1. Left: Measured fractional offset variations of the frequency ratio between three Cs fountains (FO1, FO2 and FOM as identified by the legend) and various H-masers. Long-term results span from the 2/7/2003 to the 2/11/2011, which is 3,045 days (8 years and 4 months). Right measurement between the FO2 Rb fountain and various H-masers over nearly a five year span.

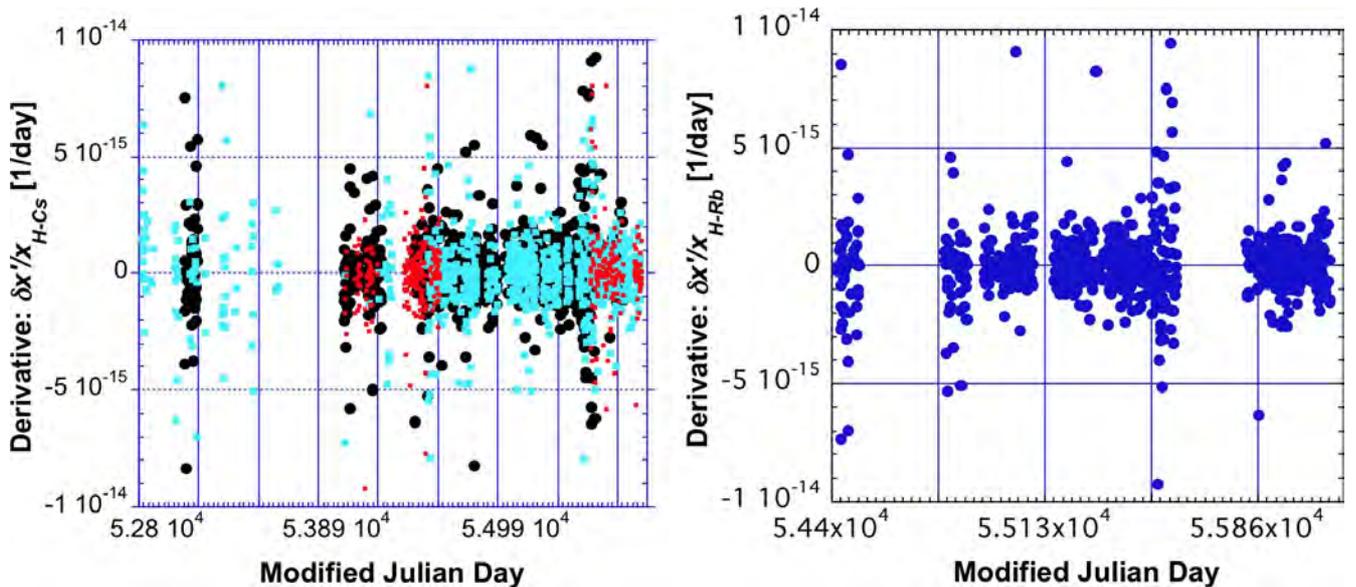

Figure 2. Left: The derivative with respect to time of the Cs versus H data shown in figure 1, in units fractional frequency per day. Each frequency measurement is averaged over 95,000 sec (1.1 days) before the derivative is taken. Right: Rb versus H data after following the same procedures. This data is used for searching for variations at the annual period.

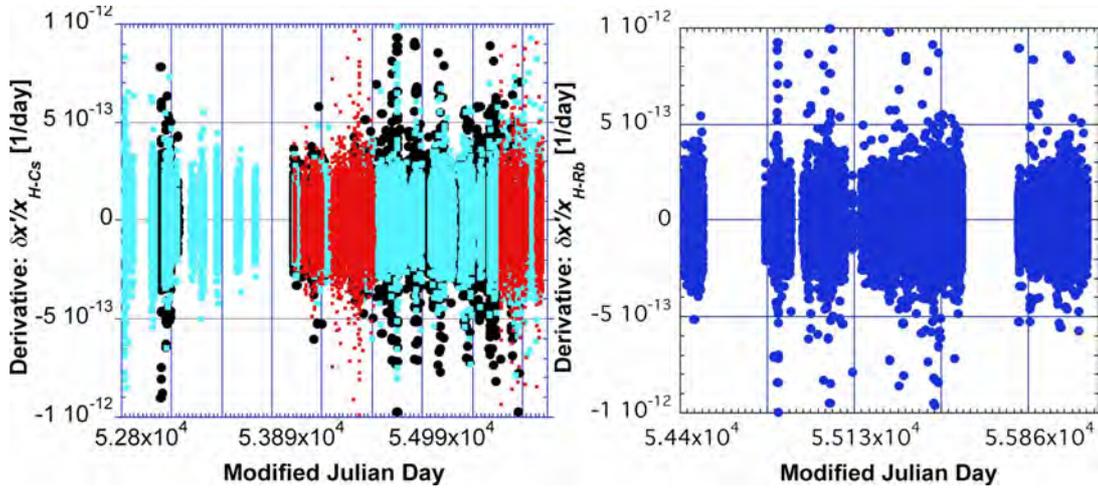

Figure 3. The derivative with respect to time of the data shown in figure 1, in units fractional frequency per day (Left: Cs versus H; Right: Rb versus H). Each frequency measurement is averaged over 2,500 sec (42 minutes) before the derivative is taken. This data is used for searching for variations at the sidereal period.

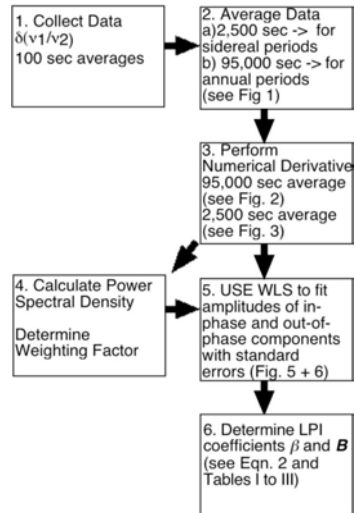

Figure 4. Flow chart, which represents the six main steps of the data collection and analysis.

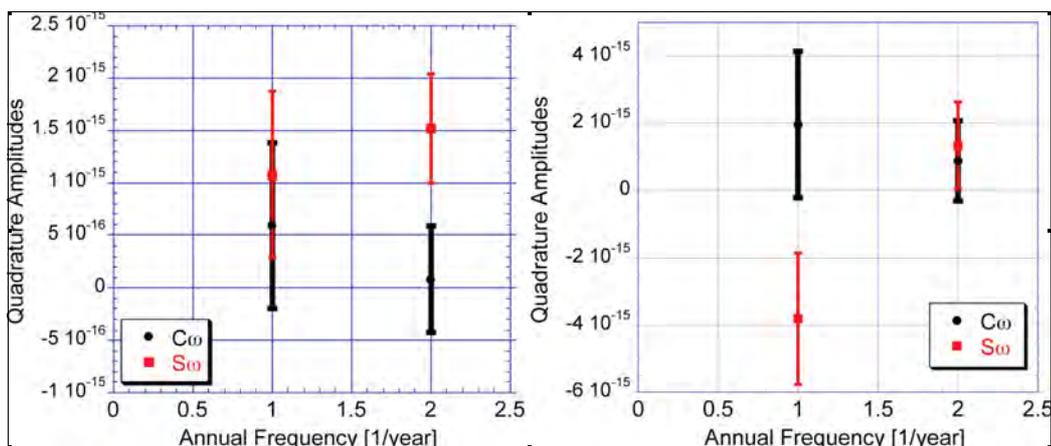

Figure 5. Results using weighted least squares to search the data from Fig 2 for the quadrature amplitudes in phase $C_\omega$ and in quadrature phase $S_\omega$ (showing the standard error as an error bar) at the annual and twice the annual frequency (Left: Cs versus H; Right: Rb versus H). Quadratures are set with respect to the phase of the perihelion (closest point to the sun) equal to zero. All results at the annual frequency are within two standard errors away from zero. Only the out of phase component for the Cs-H comparison at twice the annuual frequency has an amplitude graeter, which is a bit less than three standard errors.

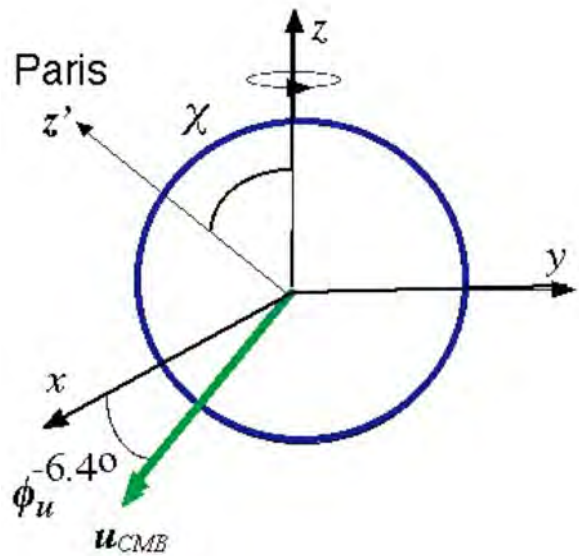

Figure 6. The Earth-centered frame[1 43 44] has the spin axis aligned with the z-axis. The velocity of the Sun with respect to the CMB, $u_{CMB}$, is defined to have no component in the y direction of the Earth frame, and is known to a precision of a few percent accuracy[46]. The Earth is spinning at the sidereal rate within this frame. The angle $\chi$ of Paris (where the experiment takes place) is shown with respect to the z-axis.

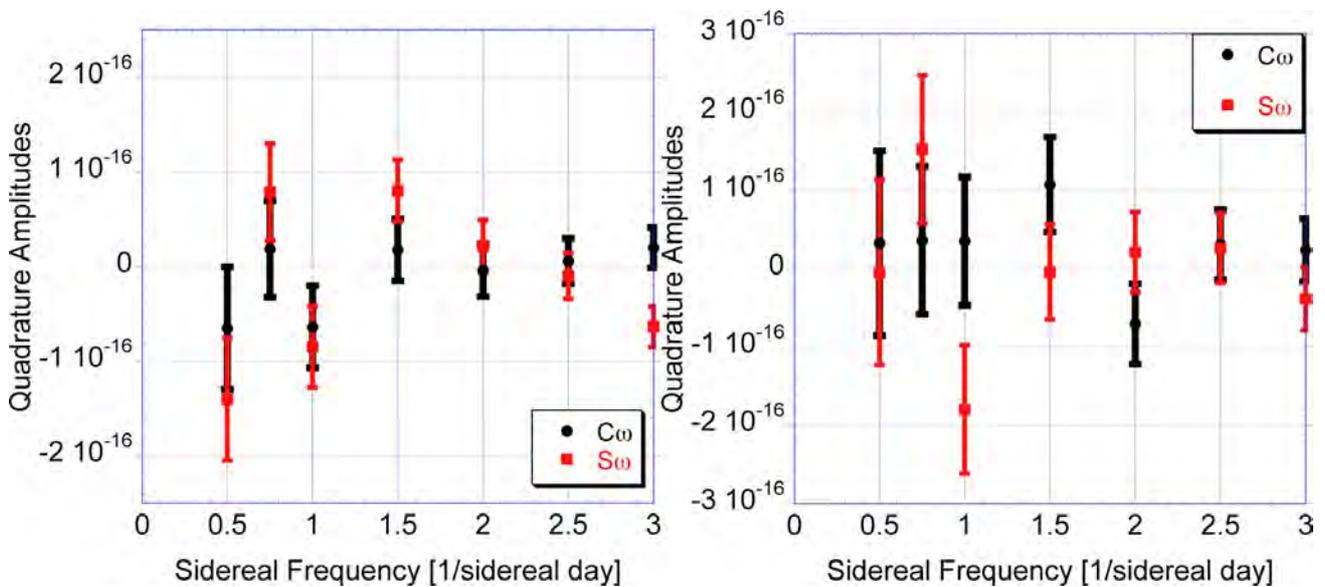

Figure 7. Calculation of boost quadratures using weighted least squares from the data in fig. 3, for multiples of the sidereal frequency for the in phase $C\omega$ and in quadrature phase $S\omega$. (Left: Cs versus H; Right: Rb versus H)